\documentclass[iop]{emulateapj}
\usepackage{apjfonts}
\makeatletter

\usepackage{graphicx}
\usepackage{amssymb}
\usepackage{amsmath}
\usepackage{hyperref}
\def\cxo{{\sl CXO\ }}

\def\gtrsim{\mathrel{\hbox{\rlap{\hbox{\lower4pt\hbox{$\sim$}}}\hbox{$>$}}}}
\def\lesssim{\mathrel{\hbox{\rlap{\hbox{\lower4pt\hbox{$\sim$}}}\hbox{$<$}}}}
\def\gtrsim{\mathrel{\hbox{\rlap{\hbox{\lower4pt\hbox{$\sim$}}}\hbox{$>$}}}}
\def\farcs{\hbox{$.\!\!^{\prime\prime}$}}

\slugcomment{}
\shorttitle{The dynamic X-ray nebula powered by the pulsar B1259--63}

\makeatother
\begin{document}

\title{The dynamic X-ray nebula powered by the pulsar B1259--63}

\author{Oleg Kargaltsev}
\affil{George Washington University,  Washington, DC, USA}
\author{George G. Pavlov}
\affil{Pennsylvania State University, Department of Astronomy \& Astrophysics, University Park, PA, USA}
\author{Martin Durant}
\affil{University of Toronto, Department of Medical Biophysics, Toronto, ON, Canada}
\author{Igor Volkov}
\affil{George Washington University,  Washington, DC, USA}
\author{Jeremy Hare}
\affil{George Washington University,  Washington, DC, USA}

\begin{abstract}
We present  observations of the  eccentric $\gamma$-ray
binary B1259--63/LS\,2883 with the {\sl Chandra} X-ray Observatory. The images reveal a variable, extended
(about $4''$,  or  $\sim1000$ times the binary orbit size) structure, which appears to be moving away from
the binary with the velocity of 0.05 of the speed of light. The observed emission is interpreted as synchrotron
radiation from relativistic particles supplied by the pulsar. However, the fast motion through the circumbinary
medium would require the emitting cloud to be loaded with a large amount of baryonic matter. 
Alternatively, the extended emission can be interpreted as  
a variable extrabinary shock in the pulsar wind outflow launched near binary apastron. The
resolved variable X-ray nebula provides an opportunity to probe   pulsar winds and their interaction with
stellar winds in a previously inaccessible way.
\end{abstract}


\section{Introduction}

Recent advances in $\gamma$-ray instrumentation have led to the discovery of GeV and TeV emission from a handful of
high-mass X-ray binaries. Most of these binaries appear to be either wind accreting or colliding wind binaries in
which a compact object (a neutron star or a black hole) is orbiting an early-B or late-O type massive
companion \citep{2013A&ARv..21...64D}.  However, for most of these high-mass $\gamma$-ray binaries (HMGBs) the energy
release mechanism and the type of compact object are unknown.   It is possible that systems with a young pulsar orbiting a massive star may dominate the population of HMGBs 
 \citep{ 2013A&ARv..21...64D}.  In these HMGBs most of the high energy emission is believed to be produced at the shock located
between the pulsar wind and the massive-star wind. Although the spatially-integrated multiwavelength spectra and
variability have been the  subject of multiple studies \citep{2006A&A...456..801D,2008A&A...477..691D,2010MNRAS.403.1873Z, 2013A&A...551A..17Z,2013MNRAS.430.2951B},
they only provide indirect diagnostics of the complex wind interaction. Theoretical models suggest that the shocked
flow can be accelerated to relativistic velocities, forming a spiral which might then be disrupted by 
instabilities \citep{2006A&A...456..801D,2008MNRAS.387...63B,2011A&A...535A..20B,2012A&A...544A..59B}. These
predictions are yet to be verified observationally because it is challenging to achieve the high angular resolution required to directly resolve the structure of the
interacting winds in these systems. Some recent intriguing evidence of extended emission comes from VLBI
observations which indicate variable, subarcsecond-scale structures in a few known HMGBs  \citep{2013arXiv1306.2830M,2012AIPC.1505..386M}.

The only HMGB in which the compact source has been
detected as a radio pulsar \citep{1992ApJ...387L..37J} is B1259--63/LS\,2883 (hereafter B1259). The radio pulsations
are seen during a large fraction of the 3.4 yr orbital period, but they disappear for about four
weeks \citep{2005MNRAS.358.1069J} when the pulsar is moving through the dense wind of its massive companion around
periastron.
  It is possible that  other HMGBs, whose orbits are more compact, also host young
pulsars which cannot be seen in the radio because the radio emission is blocked by the dense stellar wind. Studying
the high-energy emission from B1259 allows one to probe the properties of both the massive-star and pulsar winds and
study the mechanisms of their interaction. In particular, the pulsar wind properties can be studied in a regime
inaccessible in observations of isolated pulsars. In this paper we report the discovery of variable extended X-ray
emission around B1259 and discuss the implications for colliding wind models of such binaries for the theory of
pulsar winds and for the nature of high-energy emission in HMGBs.

The radio pulsar PSR B1259--63 has the following properties: period $P=47.8$\,ms, characteristic age
$\tau=P/2\dot{P}=330$\,kyr, spin-down power $\dot{E}=8.3\times10^{35}$\,erg\,s$^{-1}$, magnetic field
$B=3.3\times10^{11}$ G, and distance  $d\approx 2.3$\,kpc \citep{1992ApJ...387L..37J}. 
The orbit of the B1259 binary is highly eccentric ($e=0.87$), with semimajor axis $a\approx7$\,AU, period
$P_{\rm orb}=1236.7$\,d, and inclination  $i\approx23^{\circ}$ \citep{2011ApJ...732L..11N,2013arXiv1311.0588S}. The massive
companion, a fast-rotating late O-type star with a luminosity $L_{\star}=6.3\times10^{4} L_\odot$ \citep{2011ApJ...732L..11N}, has an equatorial excretion disk and is a source of a strong stellar wind. The
disk is thought to be inclined to the orbital plane at an angle  of $\sim10^{\circ}$ \citep{1995MNRAS.275..381M}. X-ray and radio data suggest that the pulsar passes through the disk during
time intervals of $\approx17-2$ days before periastron and $\approx6-48$ days after periastron \citep{2006MNRAS.367.1201C}.

It has been known from multiple observations that the X-ray flux and spectrum of B1259 vary with the orbital
phase \citep{2009MNRAS.397.2123C}, but no X-ray pulsations with the pulsar's period have been
detected \citep{2006MNRAS.367.1201C}. This and the relatively high X-ray luminosity of B1259, compared to isolated
pulsars with similar parameters, suggest that the observed X-ray emission is mainly caused by the collision of the
pulsar wind with the massive companion wind, which produces a shock whose properties are modulated with the orbital
period in this highly eccentric binary. Although most of the X-ray emission is likely generated within the binary
and in its immediate vicinity, the shocked matter blown out from the binary can be visible at larger distances in
high-resolution observations.

Extended radio emission of 10--50 mas ($\sim 25$--120 AU) size has been detected near periastron of B1259
and interpreted as synchrotron radiation of shocked pulsar wind ejected from the binary \citep{2011ApJ...732L..10M}. Such a size is too small to be resolved in X-ray observations, which can,
however, probe the blown-out matter at larger distances.
  
The only previous high-resolution X-ray image of B1259--63 was obtained in a 28 ks {\sl Chandra} X-ray
Observatory ({\sl CXO})  observation carried out on 2009 May 14, when the pulsar was very close to
apastron. Very faint, amorphous, asymmetric extended emission (up to $4''$ south-southwest from the pulsar
position) was detected at a $4\sigma$ confidence level \citep{2011ApJ...730....2P}. This emission was
tentatively interpreted as a pulsar wind nebula (PWN) confined by a shock between the two winds and
ultimately blown out from the binary by the wind of the massive companion. Here we report the results of two deeper \cxo observations.

\section{Observations and Data reduction}
We observed B1259 with the Advanced CCD Imaging Spectrometer (ACIS) on board \cxo on 2009 May 14, 2011
December 17, and 2013 May 19 (see Table 1 for ObsIDs and MJD dates). As the first observation was already
analyzed and the results were published \citep{2011ApJ...730....2P}, we only refer to it as needed to
supplement our analysis and discussion of the results of the last two observations. Details of the binary
phases and exposure times are listed in Table 1.

The target was imaged on the front-illuminated ACIS-I3 chip, in timed exposure mode. We
used `very faint' (VFAINT) telemetry format to reduce the detector background and optimize the sensitivity
for the faint extended emission. A 1/8 subarray was used to reduce the Frame Time to 0.4\,s and mitigate
the effect of pile-up. The highest count rate (see Table 1) corresponds to the pile-up fraction of
$<2$\%\footnote{See http://cxc.harvard.edu/proposer/POG/html/chap6.html} and hence can be neglected.

For our analysis we downloaded the pipeline-produced Level 2 event files, which have been cleaned of
various artifacts. No episodes of anomalously high background rates occurred in the observations. No
randomization was applied to the event positions in the Level 2 event files we used. To further improve
the accuracy of the event positioning, the Energy-Dependent Subpixel Event Repositioning (EDSER) procedure
was applied during the pipeline processing\footnote{See http://cxc.harvard.edu/ciao4.4/why/acissubpix.html
}. Below we consider only events with energies 0.5--8\,keV in order to minimize the background
contribution.

The detector responses for spectral analysis were produced with CIAO tools, following the standard
procedure and using the calibration database CALDB 4.4.1. Spectral fitting was done for 0.5--8\,keV with
CIAO's Sherpa modeling and fitting package.

\begin{table*}[]
\caption{Spectral fit parameters for the core and extended emission in three \cxo ACIS imaging observations \label{tab:template}}\vspace{-0.3cm}
\begin{center}
\renewcommand{\tabcolsep}{0.11cm}
\begin{tabular}{lcccccccccccc}
\tableline 
ObsID	&  MJD &	$\theta$\tablenotemark{a} 	&	$\Delta t$\tablenotemark{b}	&	Exp.\tablenotemark{c}	&	Cts\tablenotemark{d}	& 
$F_{\rm obs}$\tablenotemark{e}	&	$F_{\rm corr}$\tablenotemark{f}	&	$N_H$	&	$\Gamma$	&	$\mathcal{N}$\tablenotemark{g}	&
$\mathcal{A}$\tablenotemark{h} 	&	$\chi^2/$dof \\
&  &	deg	&	days	&	ks	&	&	$10^{-14}$ cgs	&	$10^{-14}$ cgs	&	$10^{21}$ cm$^{-2}$	&	&
$10^{-4}$	&	arcsec$^2$	&\\
\tableline 
10089 & 54965& 182 &667 &25.6 & 1825 & 139(5) & 158(6) & 1.5(7) & 1.51(10) & 2.3(3) & 2.5 & 20.9/30\\
  & & & & & 61 & 2.8(13) & 3.1(13) & 1.5$^\ast$ & 1.3(5) & 0.039(16) & 22.1 & 3.1/9 \\
14205& 55912 & 169 &370 & 56.3& 6551 & 249(4) & 296(5) & 2.9(3) & 1.39(5) & 3.8(2) & 2.5 & 76.6/87\\
    & & & & & 343 & 7.5(8) & 9.1(9) & 2.9$^\ast$ & 1.49(17)& 0.13(2) & 22.1 & 21.2/16\\
14206& 56431 & 192 &  886 & 56.3 &  4162 & 137(5) & 176(7) & 3.1(3) & 1.68(6) & 3.07(2) & 2.5 & 146/169 \\
   & & & & &  144 &  3.3(5) & 3.9(6) & 3.1$^\ast$ &  1.4(3) &  0.051(14) & 12.8  & 7.1/13 \\
\tableline 
\end{tabular} 
\end{center}
\tablecomments{ For each ObsID  the upper and the lower rows correspond to  the unresolved core and extended emission, respectively.
The $1\sigma$ uncertainties are shown in parentheses  (see also Figure 6). The XSPEC extinction model {\tt phabs} was used
throughout. Fluxes and counts are in the 0.5--8\,keV range, corrected for finite aperture size for the
core emission. An asterisk indicates that the extinction column was fixed to value of the corresponding
point source fit.}
\footnotetext[1]{True anomaly counted from periastron.}
\footnotetext[2]{Days since periastron.}
\footnotetext[3]{Exposure corrected for deadtime.}
\footnotetext[4]{Total (gross) counts.}
\footnotetext[5]{Observed flux.}
\footnotetext[6]{Extinction corrected flux.}
\footnotetext[7]{Normalization in photons\,s$^{-1}$\,cm$^{-2}$\,keV$^{-1}$ at 1\,keV.}
\footnotetext[8]{Area of the extraction region.}
\renewcommand{\thefootnote}{\arabic{footnote}}
\end{table*}

\section{Data Analysis }

\subsection{Image Analysis}

\begin{figure*}
\includegraphics[width=18cm]{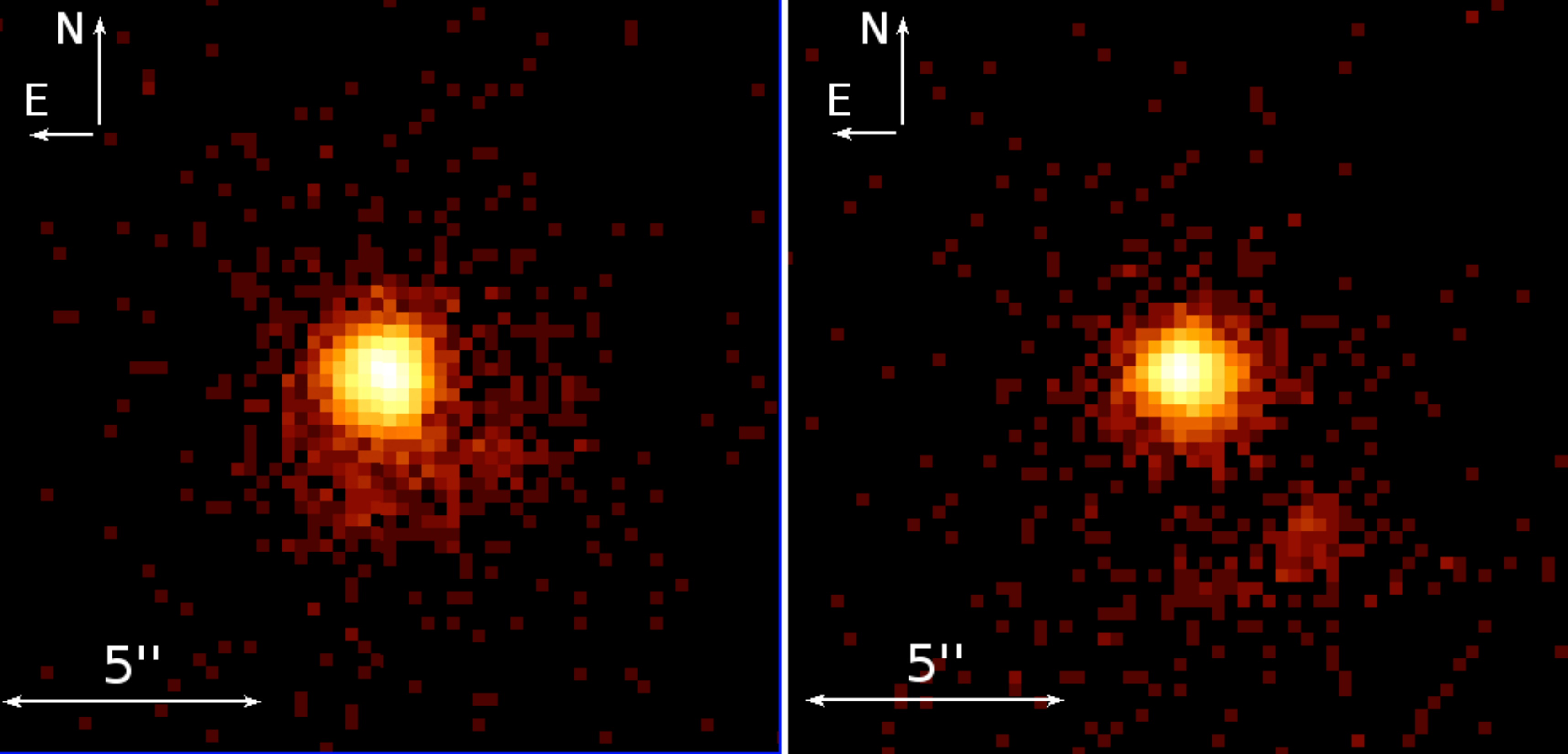}
\caption{
 \cxo ACIS images of B1259 in the 0.5--8\,keV band carried out on 2011 December 17 (left panel) and 2013
May 19 (right panel). Each pixel is 0$.\!\!^{\prime\prime}25\times0.\!\!^{\prime\prime}25$, no smoothing
or randomization is applied. See animation in \href{http://home.gwu.edu/~kargaltsev/B1259_animation.mov}{http://home.gwu.edu/$\sim$kargaltsev/B1259\_animation.mov}
\label{pic}\label{images}}
\end{figure*}

\begin{figure}
\centering 
\includegraphics[width=0.8\hsize]{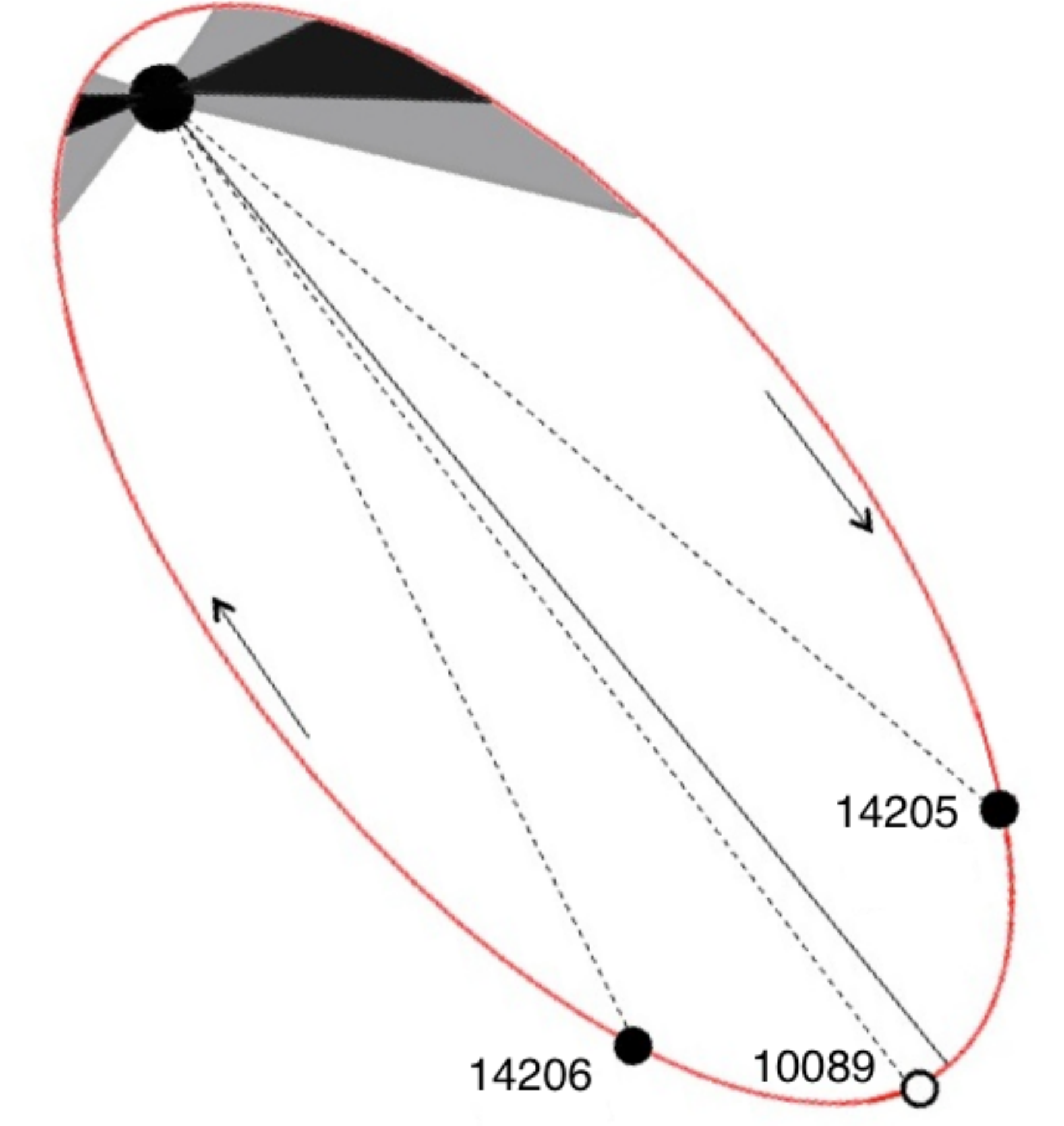}
\caption{
 Orbit of PSR B1259 around its high-mass companion. The projection on the sky shown implies that the
observed extended structure is associated with the flow launched near apastron (see text). The positions
of the pulsar at the times of the two recent CXO observations are shown by filled circles and marked by
their respective ObsIDs. The observations were carried out on 2011 December 17 and
2013 May 19, i.e., 248 days before and 268 days after the apastron passage (or 370 and 886 days after the periastron
passage), within one orbital cycle.  The orbital motion is indicated by arrows. Grey areas show the parts of the orbit
when the pulsar passes through the equatorial disk of the companion. 
\label{orb}}
\end{figure}

\begin{figure}
\centering
\includegraphics[width=0.9\hsize]{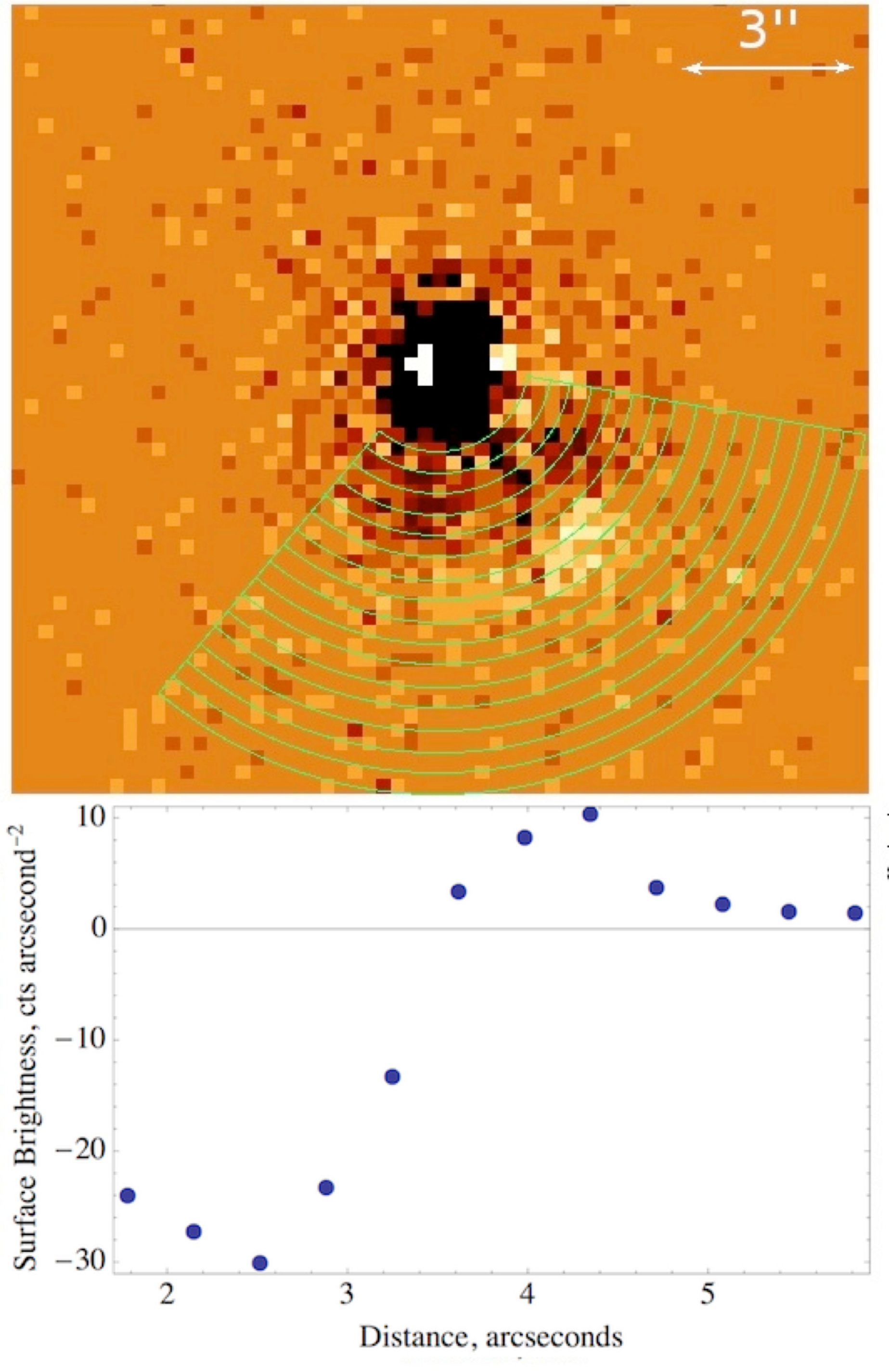}
\caption{
Top panel shows the difference of the two ACIS images shown in Figure \ref{images}. Each pixel is
0$.\!\!^{\prime\prime}25\times0.\!\!^{\prime\prime}25$, no smoothing or randomization is applied. Bottom panel shows the radial profile extracted from the difference image within the angular sector shown in the top
panel.
\label{difference}}

\end{figure}

The exquisite angular resolution of \cxo 
reveals the diffuse X-ray emission which is clearly seen to up to $5''$ from the binary (Figure \ref{images}).
Moreover, the appearance of the extended emission changed dramatically between the two observations which occurred 248 days before and 268 days after the apastron passage (see Figure \ref{orb}). While the
diffuse emission looks like a southward extension of the point-like source in the first observation, it is clearly
detached from the compact source in the second observation. This change appears as an outward motion of an arc-like
structure which is particularly well seen in the difference image shown in Figure \ref{difference}. The $1\farcs8\pm0\farcs5$ shift in the position of the arc-like structure corresponds to
the proper motion $\mu=1.27\pm0.35$ arcsec yr$^{-1} = 3.5\pm 1.0$ mas day$^{-1}$ and the projected velocity
$v_{\perp}=(0.046\pm0.013)c$ at the distance of 2.3 kpc. If the extended emission
detected in the two observations were emitted by the same object moving with a constant velocity, this object left
the binary $1260\pm 350$ days before the second observation, i.e., somewhere around the apastron passage in the
previous binary cycle.

\subsection{Spectral analysis}

\begin{figure}
\centering{\includegraphics[width=0.99\hsize]{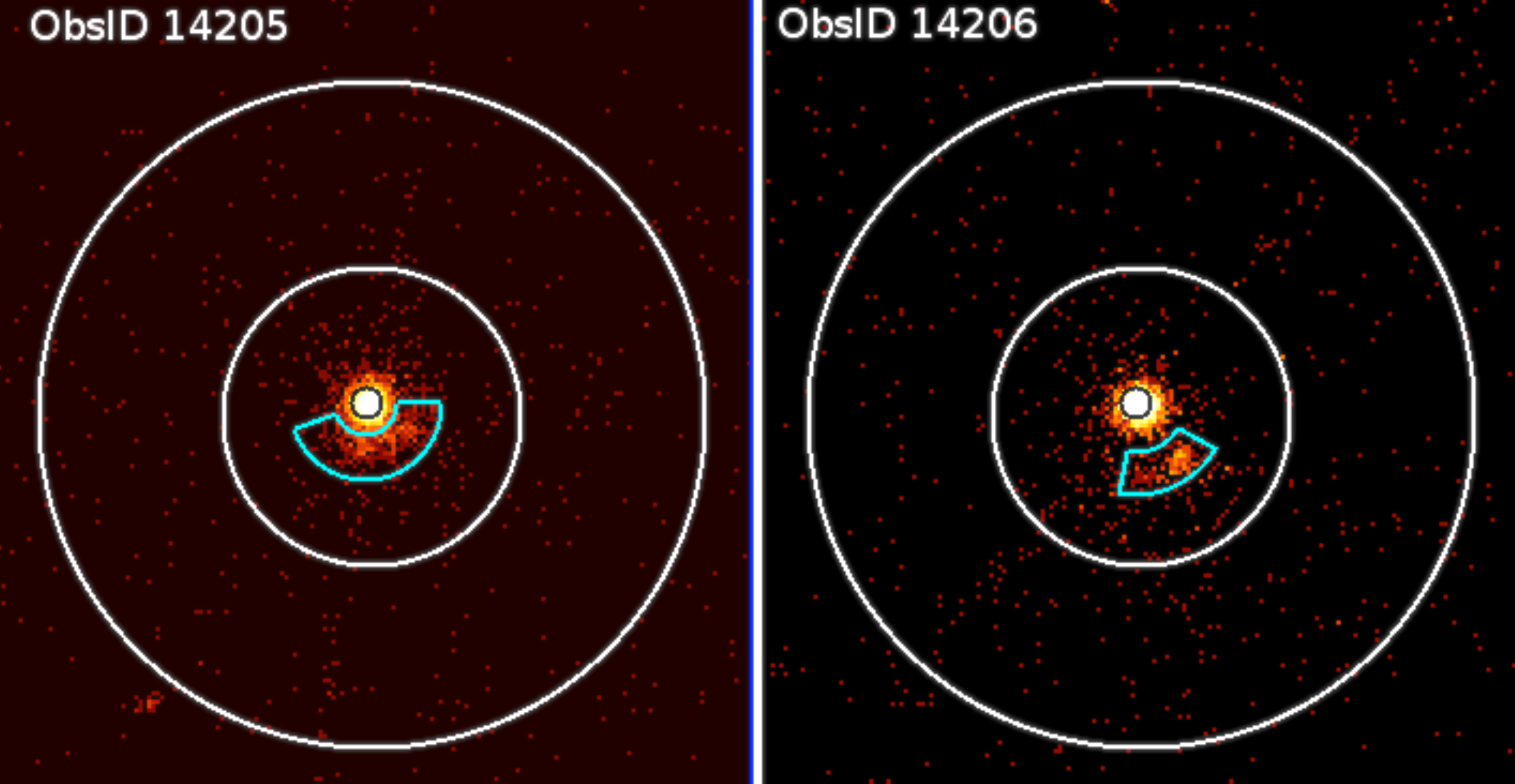}}
\caption{
The spectral extraction regions (core, extended emission, and background are shown in black, cyan, and
white, respectively) shown on top of 2011 (left) and 2013 (right)  ACIS images. Binning is
0$.\!\!^{\prime\prime}246\times0.\!\!^{\prime\prime}246$, and only 0.5--8\,keV energy photons are shown.
North is up, East is to the left. The same regions as those shown in the left panel were used to extract
the spectral properties from the 2009 observations (the corresponding image is shown in  \citealt{2011ApJ...730....2P}). The
radii of the circles are $0\farcs9$, $8\farcs55$, and $19\farcs18$.
\label{specs}}
\end{figure}

\begin{figure}
\centering
\includegraphics[width=0.99\hsize]{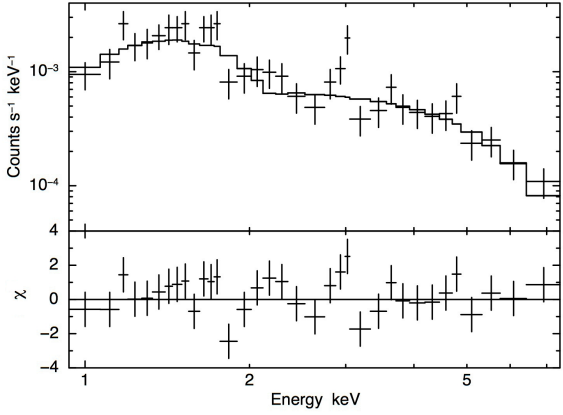}
\caption{
Spectral fits to the extended emission from December 2011 (ObsID 14205) with an absorbed PL model. See
Table 1 for the values of fitting parameters.
\label{2011spec}}
\end{figure}

The regions used for spectral extraction in ObsIDs 14205 and 14206 are shown in Figure \ref{specs}. For
completeness we also re-extracted the ObsID 10089 spectra using exactly the same extraction regions as for
ObsIDs 14205. We then binned the spectra and fit each of them to an absorbed power-law (PL) model with the
interstellar absorption cross sections from  \cite{1983ApJ...270..119M}. For each observation the extended
emission spectra were fit with the extinction column fixed at the value found for the the core component (within $r<0.9''$; includes the pulsar and unresolved PWN) in the
same observation.  An example of the extended emission spectrum (from ObsID 14205) is shown in Figure \ref{2011spec}.

\begin{figure}
\centering
\includegraphics[width=0.99\hsize]{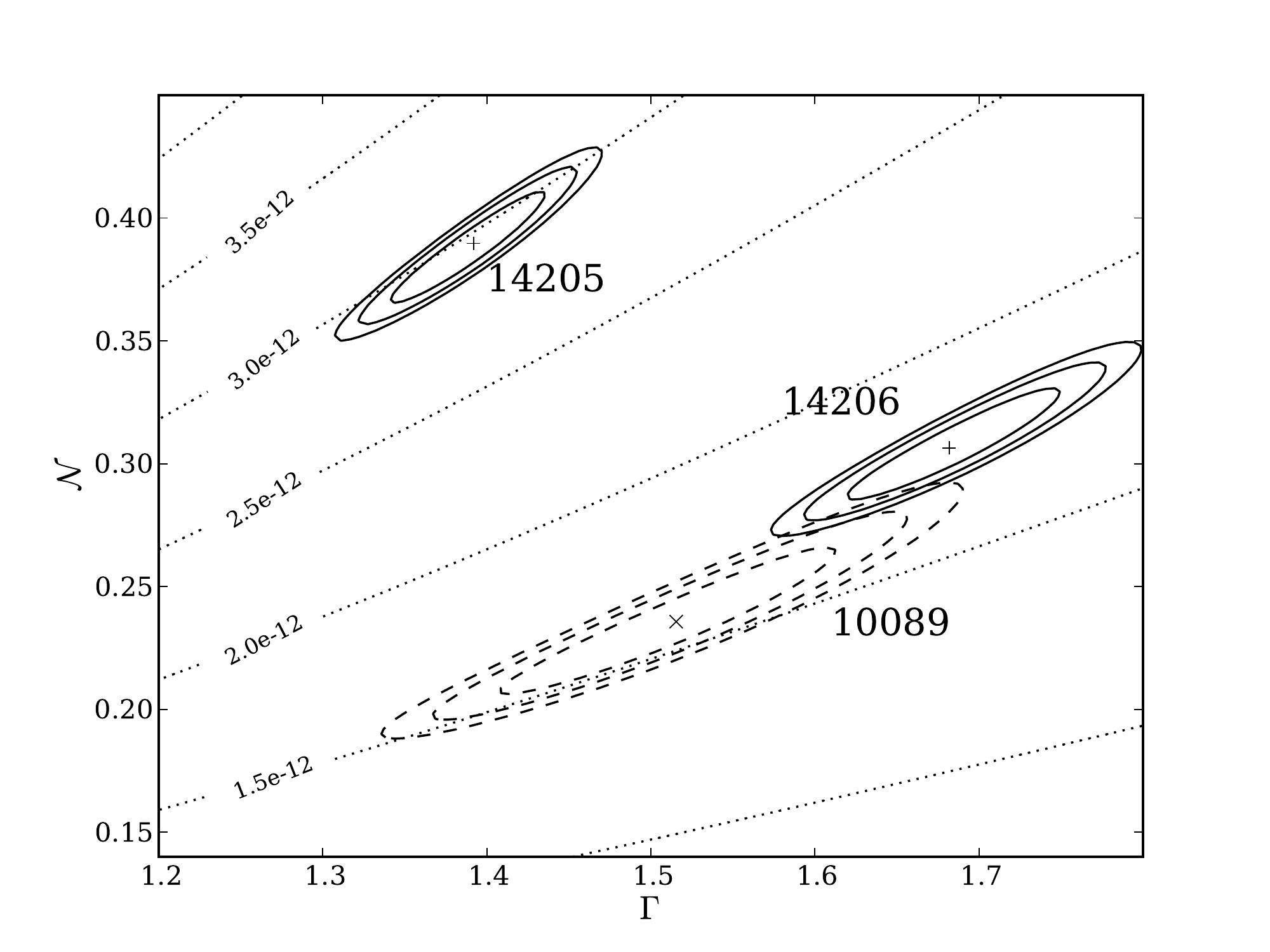}
\includegraphics[width=0.99\hsize]{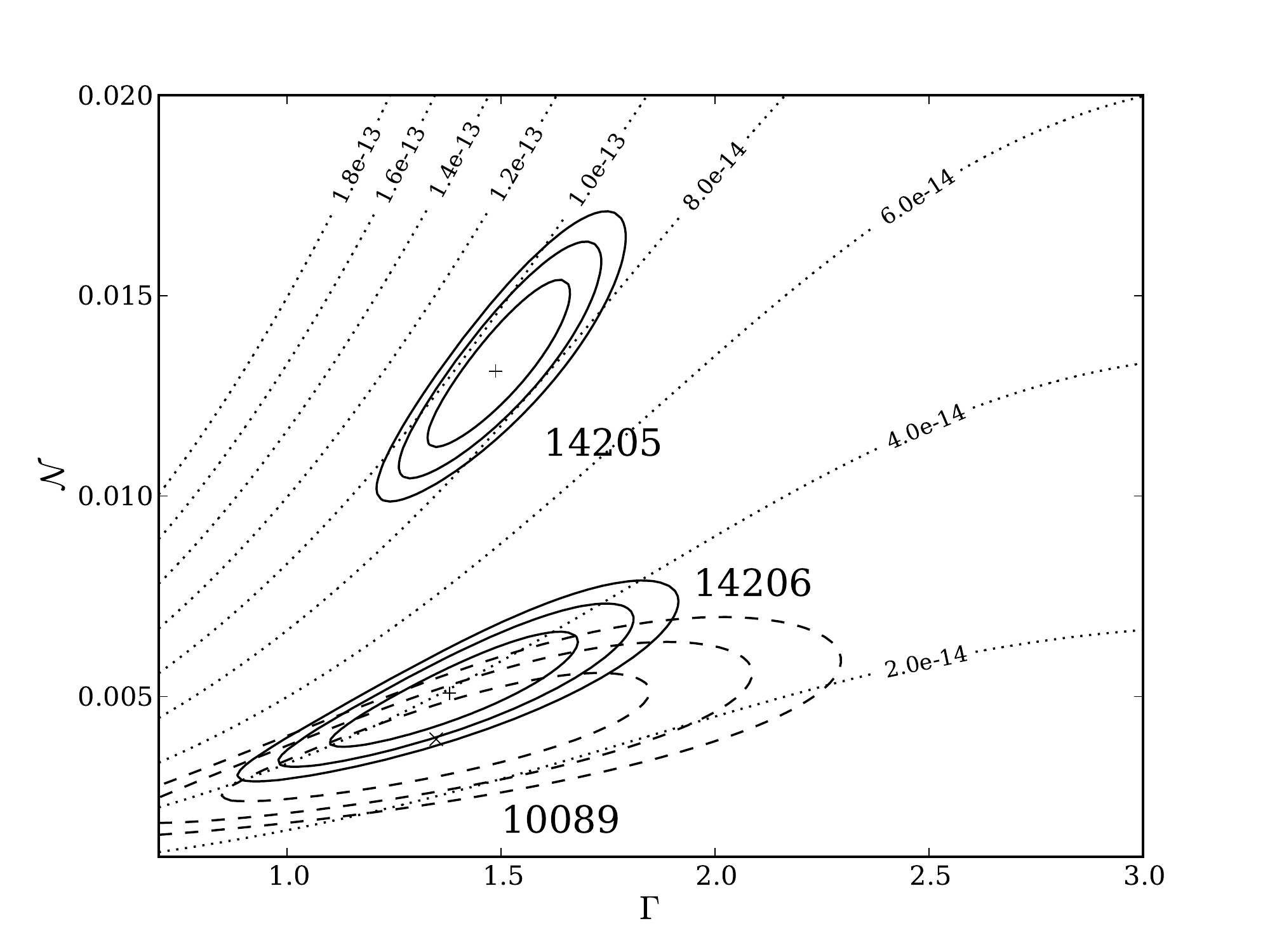}
\caption{ 
Confidence contours (68\%, 90\%, and 99\%) in the $\Gamma$--${\cal N}$ planes for the ACIS spectra of the
unresolved core ({\sl top}) and extended nebula ({\sl bottom}; see Figure \ref{specs} for region
definitions) computed for the absorbed PL model. The PL normalization ${\cal N}$ is in units of $10^{-3}$
photons cm$^{-2}$ s$^{-1}$ keV$^{-1}$ at 1 keV. The dashed lines are lines of constant unabsorbed flux in
the 0.5--8 keV band (labeled in erg\,cm$^{-2}$\,s$^{-1}$).
\label{fonts}\label{cont}}
\end{figure}

 The fitting  parameters  the two observations are presented in  Table 1 and Figure \ref{cont}.
The best-fit  photon indices,  $\Gamma\approx  1.4$--1.5,  of the extended emission  are 
similar to those obtained  for PWNe around isolated pulsars \citep{2008AIPC..983..171K}. The quality of the fit  to the
spectrum  of the extended emission from the 2011 observation (see Fig.~\ref{2011spec}) is
not perfect ($\chi_{\nu}^{2}=1.3$ for 16 degrees of freedom), but this can be explained by the noise
fluctuations due to small number of counts. The 0.5--8 keV luminosities of the extended structures, on the order of $10^{31}$ erg s$^{-1}$, are
small fractions (2\%--3\%) of the compact source luminosity.

The brightness of the core and
extended emission components,  and the  spectral slope for the core emission show appreciable variations.
The absorbed flux of the extended emission was brighter by a
factor of $\approx1.8$  in the 2011 observation (ObsID 14205) compared to the 2009 and 2013
observations
while the PL slope of the core component  became steeper by $\Delta\Gamma\simeq0.3$ in 2013 compared to 2011. At the same time no statistically significant spectral changes are measured for the
extended emission in all three observations. The limited number of counts also does not allow us to say
whether the spectrum of the extended emission differs from that of the core component. The changes in the
core flux agree with those reported from multi-year observations at low angular resolution
 \citep{2009MNRAS.397.2123C}.

The spectra of the extended emission also fit the thermal plasma model ({\tt vmekal} ), with  high plasma temperatures (best-fit $kT= 27$ keV, the upper boundary is
unconstrained, the $3\sigma$ lower limit is $\approx 6$ keV), for which the plasma radiation is dominated by
thermal bremsstrahlung. The {\tt vmekal} normalization of $3.8_{-0.5}^{+1.2}\times10^{-5}$ cm$^{-5}$
corresponds to the  number density of $\sim140 d_{2.3}^{-1/2}$ cm$^{-3}$, assuming solar
abundances and the emitting volume $V\simeq1''\times1''\times4''=2\times10^{50}d_{2.3}^3$ cm$^{3}$, where $d_{2.3}\equiv d/(2.3~{\rm kpc})$. 
 The quality of these fits is similar to that of the PL fits.

\section{Discussion}

The observed extended X-ray emission is clearly asymmetric with respect to the binary while the resolved size and separation from the binary greatly exceed the size of the binary orbit.  
It does not resemble the flow structures expected in such systems  based on the current models  \citep{2006A&A...456..801D,2008MNRAS.387...63B,2011A&A...535A..20B,2012A&A...544A..59B}.  

The location and the shape of the intrabinary shock depend on the momentum fluxes of the winds and their
angular distributions. As we know from observations of PWNe of isolated pulsars, the pulsar wind is
predominantly concentrated around the pulsar's equatorial plane. If this plane coincides with the orbital
plane, and the high-mass companion wind is nearly isotropic, the location of the contact discontinuity,
separating the shocked winds on the line connecting the two stars (the stagnation point) is at the
distance $r_{\rm in} = l \eta^{1/2}(1+\eta^{1/2})^{-1}$ from the pulsar, where $l$ is the separation
between the binary companions, $\eta = \dot{E}(\dot{M}v_w c\,\sin\Theta)^{-1}$ is the ratio of the
momentum fluxes of the pulsar wind and stellar wind at the half-distance between the pulsar and the star, $\dot{M}$ is the mass-loss rate of the high-mass companion, $v_{w}$ is terminal wind velocity, 
and $\Theta$ is the half-opening angle around the equatorial plane into which the pulsar's wind is
blowing ($\Theta=\pi/2$ for an isotropic wind). At $\eta \ll 1$ the stellar wind wraps around the pulsar and collimates the pulsar wind into a
diverging tail directed away from the high mass companion, creating a spiral pattern due to the pulsar's
orbital motion  \citep{2006A&A...456..801D}, which mixes with the denser stellar wind at the contact
discontinuity via Kelvin-Helmholtz instabilities  \citep{2011A&A...535A..20B}. At $\eta \gg 1$ the pulsar
wind takes over and collimates the stellar wind within a shock of a complex shape. According to  \cite{2005A&A...441..735M} and \cite{2008A&ARv..16..209P}, late-type O stars with $L_{\ast}<10^{5.3}L_{\odot}$
produce ``weak'' winds, with $\dot{M}=(10^{-9.5}-10^{-8})M_{\odot}$ yr$^{-1}$ and $v_w\simeq1000$--$2000$
km s$^{-1}$. Therefore, for the nearly isotropic polar component of the LS 2883 wind, the parameter $\eta
= 0.9 \dot{M}_{-8.5}^{-1} (v_{w}/1500\,{\rm km~s^{-1}})^{-1} (\sin\Theta)^{-1} $ is on the order of unity
or even larger.

The PL  spectrum of the resolved extended emission could be interpreted as synchrotron emission of the shocked pulsar wind, while the
thermal plasma emission could be produced by the massive companion wind heated to X-ray temperatures by the
interaction with the energetic pulsar wind (e.g., in oblique shocks and instabilities; \citealt{2011A&A...535A..20B,2012A&A...544A..59B}) and ejected from the system. The latter possibility,
however, requires a very large mass of the ejected plasma cloud, $m_{\rm ej}\sim  m_H n_p V \sim 5\times10^{28}
d_{2.3}^{5/2}$ g, where $n_p\sim n_e \sim 140\, d_{2.3}^{-1/2}$ cm$^{-3}$ is the proton number density estimated
from the spectral fit (see Section 3.2).
 This mass is much larger than $\dot{M} P_{\rm orb}=6.7\times 10^{25} \dot{M}_{-8}$ g lost by
the high-mass companion wind during the whole orbital period. Therefore, we conclude that thermal emission
from the heated O-star ejecta alone cannot explain the observed extended X-ray emission. 

The other possible interpretation could be synchrotron 
emission from pulsar wind particles.  Using standard formulae for synchtrotron radiation (e.g.,  \citealt{1965ARA&A...3..297G,1970ranp.book.....P}),
we can estimate the Lorentz factors of the emitting electrons as $\gamma \sim (10^7 - 10^8)
(B/100\,\mu{\rm G})^{1/2}$. One can also express the magnetic field in a synchrotron-emitting region in
terms of observable parameters  \citep{2003ApJ...591.1157P}:
\begin{equation}
B=27\,\left\{\frac{k_m
}{a_p (3-2\Gamma)}
\left[E_{{\rm M},p}^{(3-2\Gamma)/2} - E_{{\rm m},p}^{(3-2\Gamma)/2}\right]
\frac{\mathcal{B}_{-7}}{\bar{s}_{16}}\right\}^{2/7}\,\,\mu{\rm G}\,.
\end{equation}
Here $k_m$ is the ratio of the magnetic field energy density to the energy density of the relativistic
particles, $\mathcal{B}=\mathcal{N}/\mathcal{A} = 10^{-7} \mathcal{B}_{-7}$ photons
(s\,cm$^2$\,keV\,arcsec$^2$)$^{-1}$ is the average spectral surface brightness at $E=1$ keV (see Table 1), $\mathcal{N}$ is the normalization of photon spectral flux measured from the area $\mathcal{A}$
(in arcsec$^2$), $\bar{s}=10^{16}\, \bar{s}_{16}$ cm is the average length of the radiating region along
the line of sight, $E_{{\rm m},p}=E_{\rm m}/y_{{\rm m},p}$, $E_{{\rm M},p}=E_{\rm M}/y_{{\rm M},p}$,
$E_{\rm m}$ and $E_{\rm M}$ are the lower and upper energies of the photon PL spectrum (in keV), and
$y_{{\rm m},p}$, $y_{{\rm M},p}$ and $a_p$ are the numerical coefficients whose values depend on the slope
$p=2\Gamma-1$ of the electron PL spectral energy distribution (\citealt{1965ARA&A...3..297G}). Taking $\Gamma$, $\mathcal{N}$, and $\mathcal{A}$ values from Table 1,
and assuming $\bar{s}_{16}=3.4$ (equivalent to $1''$ at $d=2.3$ kpc), we obtain $B\sim80k_m^{2/7}
d_{2.3}^{-2/7}$~$\mu$G for the extended emission seen in the 2011 and 2013 observations (for $E_{\rm m}=0.5$
keV, $E_{\rm M}=8$ keV). Note that for the measured $\Gamma=1.4$--$1.5$ the magnetic field dependence on the
(unknown) values of $E_{\rm m}$ and $E_{\rm M}$ is extremely weak (e.g., expanding the energy range by
five orders of magnitude up and down increases the magnetic field value by a factor of 1.6). The lower
limit on magnetic field can be obtained by requiring the electron Larmor radius to be smaller than the
transverse size ($\sim 1''$) of the arc-like extended emission seen in the 2013 image, $B\gtrsim 11
d_{2.3}^{-2/3}(E_{\rm syn}/3~{\rm keV})^{-1/3}$~$\mu$G (which implies $k_m\gtrsim10^{-4}$). The
corresponding magnetic and electron energies are $W_m \sim 5\times10^{40} k_m^{4/7} d_{2.3}^{17/7}$ erg
and $W_e\sim 5\times10^{40} k_m^{-3/7} d_{2.3}^{17/7}$ erg, respectively, in the volume $V=2\times10^{50}
d_{2.3}^3$ cm$^{3}$. The total energy, $W=W_m+W_e$, in the emitting volume is much smaller than the energy
lost by the pulsar spin-down during one orbital period, $\dot{E}P_{\rm orb} =8.8\times10^{43}~{\rm erg}$,
in a broad range of magnetization parameter, $3\times 10^{-8} d_{2.3}^{17/3} \ll k_m \ll 5\times 10^5
d_{2.3}^{-17/4}$, which means that the cloud energy can be supplied by the pulsar in a fraction of the
binary period.  In the following  subsections we describe two scenarios that could explain the origin of the extended synchrotron-emitting structure.
 

\subsection{Fast moving cloud ejected from the binary.}

The first scenario  assumes  that we observed the
same cloud in December 2011 and May 2013, which had been moving with the average velocity $v\sim 0.05 c$ since the
launch. However, if such a cloud is created and ejected from the binary in the course of interaction of the pulsar
wind with the dense excretion disk close to the nearest preceding periastron (i.e., about one year before the first
observation), it should decelerate very rapidly. The drag force exerted on the light, fast-moving cloud by the
ambient medium is very large: $f\sim \rho_{\rm amb} v^2 A$, where $A\sim 5\times 10^{33} d_{2.3}^2$ cm$^2$ is the
frontal area of the cloud, and $\rho_{\rm amb}\sim n_H m_H$ is the ambient density. The deceleration caused by this
force can be estimated as $|\dot{v}|\sim f c^2 W^{-1} \sim 3\times 10^8 n_{H} (v/0.05 c)^2 (k_m^{4/7} +
k_m^{-3/7})^{-1} d_{2.3}^{-3/7}$ cm s$^{-2}$, where $W\sim 5\times10^{40} (k_m^{4/7} + k_m^{-3/7}) d_{2.3}^{17/7}$
erg is the total (particle and magnetic field) energy in the emitting volume.
The characteristic deceleration time is then $t_{\rm dec} \sim W v f^{-1} c^{-2} \sim 4 n_H^{-1} (v/0.05 c)^{-1}
(k_m^{4/7} + k_m^{-3/7}) d_{2.3}^{3/7}$ s. This means that if the circumbinary medium is filled with the stellar
wind of the high-mass companion, the cloud of relativistic electrons would be decelerated and/or destroyed in a time
much shorter than the interval of 519 days between our two observations. The deceleration time could be larger if
the cloud is loaded with a substantial amount of the O-star wind. For instance, it will exceed the apparent travel
time from the binary, $t_{\rm trav}\sim 1260$ d, if $m_{\rm load} > 1.3\times 10^{27} n_H (v/0.05 c)(t_{\rm
trav}/1260\,{\rm d}) d_{2.3}^2$ g, which still requires a rather low ambient density, $n_H\ll 10^{-2}$ cm$^{-3}$, to
be consistent with the mass and energy budget. Therefore, this scenario seems rather unlikely. The drag force could
be much smaller if the matter ahead of the cloud is mostly comprised of the pulsar wind particles and is moving away
from the binary with a high velocity. The presence of such a lepton-dominated medium would be more plausible on the
apastron side of the orbit where the pulsar spends most of the time. Overall, the moving cloud scenario seems to
require somewhat artificial assumptions, but it cannot be firmly ruled out without further investigation (e.g.,
another {\sl CXO} observation could confirm or rule out the translational motion of the cloud).

\subsection{Varibale extrabinary termination shock}

In the second scenario, which we consider more plausible, the extended structures are formed at the {\em extrabinary
termination shock} in the pulsar wind at large distances from the binary, comparable to those observed in PWNe
around isolated pulsars \citep{2008AIPC..983..171K}, so that the different locations of the structures in December
2011 and May 2013 are due to different conditions at the epochs of their formation, rather than to the motion of a
cloud of relativistic electrons from the binary.

Formation of such a distant termination shock can be understood as follows. In a high mass binary the pulsar's wind
collides with the companion's wind and forms a double-shock structure within the binary and in its
vicinity \citep{1997ApJ...477..439T}. If the pulsar wind is dynamically dominant, it escapes freely in the directions away from the high-mass companion and, similar to an isolated pulsar
wind, shocks in the ambient medium at a distance $r_{\rm out} \sim (\dot{E}/4\pi c p_{\rm amb}\sin\Theta)^{1/2}$,
where the wind's ram pressure becomes comparable to the ambient pressure, $p_{\rm amb}$. Synchrotron radiation from the shocked pulsar wind can be observed as an extrabinary PWN. As
the pulsar orbits the heavy companion, this extrabinary shock rotates around the binary, seeding the medium with
X-ray emitting relativistic electrons (positrons). If their cooling time is larger than the orbital period, one can
expect a uniformly bright extended structure to form around a circular binary at distances $\gtrsim r_{\rm out}$;
such a structure would be brighter on the apastron side for an eccentric binary because the pulsar spends more time
around apastron than around periastron. The location of the structure can vary due to variations in the ambient
pressure or because of instabilities that can destroy the contact discontinuity, separating the two winds, and mix
the stellar wind matter into the pulsar wind.

The extended X-ray structures detected near the B1259 binary could be associated with such an extrabinary
termination shock formed at a distance $r_{\rm out} = 1.5\times 10^{17}~p_{-10}^{-1/2} (\sin\Theta)^{-1/2}$ cm,
where $p_{-10}=p_{\rm amb}/10^{-10}$ dyn cm$^{-2}$. The corresponding angular separation from the binary, $4\farcs4~
p_{-10}^{-1/2} (\sin\Theta)^{-1/2} \sin\alpha~d_{2.3}^{-1}$ (where $\alpha$ is the angle between the line of sight
and the line connecting the binary and the emitting region) is close to the observed separations of the detected
extended structures at $ p_{-10}\sim1$. Such high pressures are possible in wind-blown bubbles around massive
stars \citep{2004RMxAC..22..136V}.

 The magnetic field immediately beyond the shock can be estimated as $B_s \sim 3 (2\dot{E}\sigma/c r_{\rm
out}^2)^{1/2} \sim 150 p_{-10}^{1/2}\sigma^{1/2}$ $\mu$G for $\sigma\ll 1$, where $\sigma$ is the pre-shock
magnetization parameter \citep{1984ApJ...283..694K}. This field is consistent with $B\sim 80 k_m^{2/7}$ $\mu$G,
inferred assuming a synchrotron origin of the detected emission. The synchrotron
cooling time in such a field, $\tau_{\rm syn} \sim 10 p_{-10}^{-3/4}\sigma^{-3/4}$ yr, exceeds the orbital
period, which means that the PWN structure created in a given orbital cycle could be seen in a few following
cycles, and its azimuthal extent around the binary can be rather large. The latter conclusion does not contradict
to the small azimuthal extent ($\sim 40^\circ - 60^\circ$) of the observed structures because the PWN brightness
is proportional to the amount of pulsar wind particles supplied in a given direction. This amount strongly varies
throughout the eccentric orbit, with a maximum around apastron. It should be noted that a major restructuring of
both the extrabinary and intrabinary PWN components is expected when the pulsar moves through the dense
excretion disk of the O star around periastron. However, since the pulsar spends only a small fraction ($\sim
0.05$) of the orbital period in this stage but sweeps a large range ($>180^\circ$) of true anomalies (see Figure 
\ref{orb}), we do not expect a significant contribution of this stage to the extrabianry PWN. The observed shift of the
extended structure during 1.5 years can be reconciled with the large $\tau_{\rm syn}$ if the apparent motion is
interpreted as a shift of the termination shock caused by a 30\% change of the ratio of the pulsar-wind pressure
to the ambient pressure. 

The extended emission was the faintest during the 2009 observation, which occurred near
apastron at the true anomaly intermediate between those of the 2011 and 2013 observations. Therefore, the
extended emission flux is not a periodic function of  true anomaly, even for adjacent orbital cycles, which
could be explained either by  variations in the density/velocity of the O-star wind or by instabilities
developing at the interface between the winds. These arguments may also explain a factor of three different X-ray
fluxes from the entire binary (at nearly the same true anomalies) seen in the earlier observations with other
satellites (see Figure 2 from  \citealt{2009MNRAS.397.2123C}). In this interpretation we neglected the role of possible mixing of the
stellar wind into the pulsar wind, which could lead to mass loading and deceleration of the pulsar wind at large
distances from the binary.

\section{Conclusions}

The unexpected morphology and transient behavior of the resolved extended emission have not been captured in the
existing models of such systems. Most likely, the observed structures can be interpreted as the synchrotron emission
from the pulsar wind  shocked by the interaction with the circumbinary medium or with the heavy stellar wind
entrained into the pulsar wind outflow. The appearance of the PWN and the  parameter estimates suggest that
in either case most of the observed extended emission comes from the pulsar wind ejected from the binary near
apastron. If our interpretation of the discovered X-ray emission is correct, then B1259 can be used to probe the
properties of  stellar and pulsar winds, to study baryonic mass loading of pulsar winds, and to shed light on the
complex relativistic shock dynamics. Systematic monitoring of B1259, using high-resolution X-ray,
UV, optical, IR, and mm-band observations, is needed to fully utilize this unique laboratory offered by nature.

\medskip\noindent{\bf Acknowledgments:}
This work was supported by National Aeronautics Space Administration grants  NNX09AC81G and 
NNX09AC84G, and through {\sl CXO} Award GO2-13085 issued by the {\sl CXO} X-ray Observatory Center, which is operated by the
Smithsonian Astrophysical Observatory for and on behalf of the National Aeronautics Space Administration
under contract NAS8-03060.


\begin{thebibliography}{}

\bibitem[\protect\citeauthoryear{{Bednarek} \& {Sitarek}}{{Bednarek} \&
  {Sitarek}}{2013}]{2013MNRAS.430.2951B}
{Bednarek}, W.,  \& {Sitarek}, J. 2013, \mnras, 430, 2951

\bibitem[\protect\citeauthoryear{{Bogovalov} et~al.}{{Bogovalov}
  et~al.}{2008}]{2008MNRAS.387...63B}
{Bogovalov}, S.~V., {Khangulyan}, D.~V., {Koldoba}, A.~V., {Ustyugova}, G.~V.,
  \& {Aharonian}, F.~A. 2008, \mnras, 387, 63

\bibitem[\protect\citeauthoryear{{Bosch-Ramon} \& {Barkov}}{{Bosch-Ramon} \&
  {Barkov}}{2011}]{2011A&A...535A..20B}
{Bosch-Ramon}, V.,  \& {Barkov}, M.~V. 2011, \aap, 535, A20

\bibitem[\protect\citeauthoryear{{Bosch-Ramon} et~al.}{{Bosch-Ramon}
  et~al.}{2012}]{2012A&A...544A..59B}
{Bosch-Ramon}, V., {Barkov}, M.~V., {Khangulyan}, D.,  \& {Perucho}, M. 2012,
  \aap, 544, A59

\bibitem[\protect\citeauthoryear{{Chernyakova} et~al.}{{Chernyakova}
  et~al.}{2009}]{2009MNRAS.397.2123C}
{Chernyakova}, M., {Neronov}, A., {Aharonian}, F., {Uchiyama}, Y.,  \&
  {Takahashi}, T. 2009, \mnras, 397, 2123

\bibitem[\protect\citeauthoryear{{Chernyakova} et~al.}{{Chernyakova}
  et~al.}{2006}]{2006MNRAS.367.1201C}
{Chernyakova}, M., {Neronov}, A., {Lutovinov}, A., {Rodriguez}, J.,  \&
  {Johnston}, S. 2006, MNRAS, 367, 1201

\bibitem[\protect\citeauthoryear{{Dubus}}{{Dubus}}{2006}]{2006A&A...456..801D}
{Dubus}, G. 2006, \aap, 456, 801

\bibitem[\protect\citeauthoryear{{Dubus}}{{Dubus}}{2013}]{2013A&ARv..21...64D}
{Dubus}, G. 2013, \aapr, 21, 64

\bibitem[\protect\citeauthoryear{{Dubus}, {Cerutti}, \& {Henri}}{{Dubus}
  et~al.}{2008}]{2008A&A...477..691D}
{Dubus}, G., {Cerutti}, B.,  \& {Henri}, G. 2008, \aap, 477, 691

\bibitem[\protect\citeauthoryear{{Ginzburg} \& {Syrovatskii}}{{Ginzburg} \&
  {Syrovatskii}}{1965}]{1965ARA&A...3..297G}
{Ginzburg}, V.~L.,  \& {Syrovatskii}, S.~I. 1965, \araa, 3, 297

\bibitem[\protect\citeauthoryear{{Johnston} et~al.}{{Johnston}
  et~al.}{2005}]{2005MNRAS.358.1069J}
{Johnston}, S., {Ball}, L., {Wang}, N.,  \& {Manchester}, R.~N. 2005, \mnras,
  358, 1069

\bibitem[\protect\citeauthoryear{{Johnston} et~al.}{{Johnston}
  et~al.}{1992}]{1992ApJ...387L..37J}
{Johnston}, S., {Manchester}, R.~N., {Lyne}, A.~G., {Bailes}, M., {Kaspi},
  V.~M., {Qiao}, G.,  \& {D'Amico}, N. 1992, \apjl, 387, L37

\bibitem[\protect\citeauthoryear{{Kargaltsev} \& {Pavlov}}{{Kargaltsev} \&
  {Pavlov}}{2008}]{2008AIPC..983..171K}
{Kargaltsev}, O.,  \& {Pavlov}, G.~G. 2008, in American Institute of Physics
  Conference Series, Vol. 983, 40 Years of Pulsars: Millisecond Pulsars,
  Magnetars and More, ed. C.~{Bassa}, Z.~{Wang}, A.~{Cumming}, \& V.~M.
  {Kaspi}, 171

\bibitem[\protect\citeauthoryear{{Kennel} \& {Coroniti}}{{Kennel} \&
  {Coroniti}}{1984}]{1984ApJ...283..694K}
{Kennel}, C.~F.,  \& {Coroniti}, F.~V. 1984, \apj, 283, 694

\bibitem[\protect\citeauthoryear{{Martins} et~al.}{{Martins}
  et~al.}{2005}]{2005A&A...441..735M}
{Martins}, F., {Schaerer}, D., {Hillier}, D.~J., {Meynadier}, F.,
  {Heydari-Malayeri}, M.,  \& {Walborn}, N.~R. 2005, \aap, 441, 735

\bibitem[\protect\citeauthoryear{{Melatos}, {Johnston}, \& {Melrose}}{{Melatos}
  et~al.}{1995}]{1995MNRAS.275..381M}
{Melatos}, A., {Johnston}, S.,  \& {Melrose}, D.~B. 1995, \mnras, 275, 381

\bibitem[\protect\citeauthoryear{{Mold{\'o}n} et~al.}{{Mold{\'o}n}
  et~al.}{2011}]{2011ApJ...732L..10M}
{Mold{\'o}n}, J., {Johnston}, S., {Rib{\'o}}, M., {Paredes}, J.~M.,  \&
  {Deller}, A.~T. 2011, \apjl, 732, L10

\bibitem[\protect\citeauthoryear{{Mold{\'o}n}, {Rib{\'o}}, \&
  {Paredes}}{{Mold{\'o}n} et~al.}{2012}]{2012AIPC.1505..386M}
{Mold{\'o}n}, J., {Rib{\'o}}, M.,  \& {Paredes}, J.~M. 2012, in American
  Institute of Physics Conference Series, Vol. 1505, American Institute of
  Physics Conference Series, ed. F.~A. {Aharonian}, W.~{Hofmann}, \& F.~M.
  {Rieger}, 386

\bibitem[\protect\citeauthoryear{{Mold{\'o}n}, {Rib{\'o}}, \&
  {Paredes}}{{Mold{\'o}n} et~al.}{2013}]{2013arXiv1306.2830M}
{Mold{\'o}n}, J., {Rib{\'o}}, M.,  \& {Paredes}, J.~M. 2013, ArXiv e-prints

\bibitem[\protect\citeauthoryear{{Morrison} \& {McCammon}}{{Morrison} \&
  {McCammon}}{1983}]{1983ApJ...270..119M}
{Morrison}, R.,  \& {McCammon}, D. 1983, \apj, 270, 119

\bibitem[\protect\citeauthoryear{{Negueruela} et~al.}{{Negueruela}
  et~al.}{2011}]{2011ApJ...732L..11N}
{Negueruela}, I., {Rib{\'o}}, M., {Herrero}, A., {Lorenzo}, J., {Khangulyan},
  D.,  \& {Aharonian}, F.~A. 2011, \apjl, 732, L11

\bibitem[\protect\citeauthoryear{{Pacholczyk}}{{Pacholczyk}}{1970}]{1970ranp.book.....P}
{Pacholczyk}, A.~G. 1970, {Radio astrophysics. Nonthermal processes in galactic
  and extragalactic sources}, Series of Books in Astronomy and Astrophysics
  (San Francisco: Freeman)

\bibitem[\protect\citeauthoryear{{Pavlov}, {Chang}, \& {Kargaltsev}}{{Pavlov}
  et~al.}{2011}]{2011ApJ...730....2P}
{Pavlov}, G.~G., {Chang}, C.,  \& {Kargaltsev}, O. 2011, ApJ, 730, 2

\bibitem[\protect\citeauthoryear{{Pavlov} et~al.}{{Pavlov}
  et~al.}{2003}]{2003ApJ...591.1157P}
{Pavlov}, G.~G., {Teter}, M.~A., {Kargaltsev}, O.,  \& {Sanwal}, D. 2003, \apj,
  591, 1157

\bibitem[\protect\citeauthoryear{{Puls}, {Vink}, \& {Najarro}}{{Puls}
  et~al.}{2008}]{2008A&ARv..16..209P}
{Puls}, J., {Vink}, J.~S.,  \& {Najarro}, F. 2008, Astronomy and Astrophysics
  Review, 16, 209

\bibitem[\protect\citeauthoryear{{Shannon}, {Johnston}, \&
  {Manchester}}{{Shannon} et~al.}{2013}]{2013arXiv1311.0588S}
{Shannon}, R.~M., {Johnston}, S.,  \& {Manchester}, R.~N. 2013, ArXiv e-prints

\bibitem[\protect\citeauthoryear{{Tavani} \& {Arons}}{{Tavani} \&
  {Arons}}{1997}]{1997ApJ...477..439T}
{Tavani}, M.,  \& {Arons}, J. 1997, \apj, 477, 439

\bibitem[\protect\citeauthoryear{{van Marle}, {Langer}, \&
  {Garc{\'{\i}}a-Segura}}{{van Marle} et~al.}{2004}]{2004RMxAC..22..136V}
{van Marle}, A.~J., {Langer}, N.,  \& {Garc{\'{\i}}a-Segura}, G. 2004, in
  Revista Mexicana de Astronomia y Astrofisica Conference Series, Vol.~22,
  Gravitational Collapse: from Massive Stars to Planets, ed.
  G.~{Garcia-Segura}, G.~{Tenorio-Tagle}, J.~{Franco}, \& H.~W. {Yorke}, 136

\bibitem[\protect\citeauthoryear{{Zabalza} et~al.}{{Zabalza}
  et~al.}{2013}]{2013A&A...551A..17Z}
{Zabalza}, V., {Bosch-Ramon}, V., {Aharonian}, F.,  \& {Khangulyan}, D. 2013,
  \aap, 551, A17

\bibitem[\protect\citeauthoryear{{Zdziarski}, {Neronov}, \&
  {Chernyakova}}{{Zdziarski} et~al.}{2010}]{2010MNRAS.403.1873Z}
{Zdziarski}, A.~A., {Neronov}, A.,  \& {Chernyakova}, M. 2010, \mnras, 403,
  1873

\end{thebibliography}
\end{document}